\documentstyle[prb,preprint,aps]{revtex}
 
\begin{document}
\draft
\widetext

%
%
%
\title{Theory for Superconducting Properties of the
Cuprates:\\
Doping Dependence of the Electronic Excitations and
Shadow States}
\author{ S. Grabowski, M. Langer, J. Schmalian, and
 K. H. Bennemann}
\address{ Institut f\"ur Theoretische Physik,
 Freie Universit\"at Berlin, Arnimallee 14, 14195 Berlin ,
Germany}

\date{\today}
\maketitle

\widetext
\begin{abstract}
\leftskip 54.8pt
\rightskip 54.8pt
The superconducting phase of the 2D one-band Hubbard
model is studied within the FLEX approximation
and by using an Eliashberg theory. We investigate the
doping dependence of $T_c$, of the gap function
$\Delta ({\bf k},\omega)$ and of the effective pairing
interaction. Thus we find that $T_c$ becomes
maximal for $13 \; \%$ doping. In {\it overdoped}
systems $T_c$ decreases due to the weakening of the
antiferromagnetic correlations, while in the {\it underdoped}
 systems due to the decreasing quasi
particle lifetimes. Furthermore, we find {\it shadow states}
below $T_c$ which affect the electronic
excitation spectrum and lead to fine structure in
 photoemission experiments.
\end{abstract}



\narrowtext
There is still no definite understanding of the pairing
 interaction in the High-T$_c$ superconductors.
However, improved experimental techniques like phase
sensitive
measurements of the superconducting order
parameter~\cite{wollman} and angular resolved photoemission
(ARPES) ~\cite{shen} have revealed important new
structures in the electronic excitation spectrum that
might help to clarify the nature of superconductivity
 in the cuprates. The observation of shadows of the
Fermi surface (FS) in the normal state by
Aebi {\it et al.}~\cite{aebi} and the strong evidence for a
$d_{x^2-y^2}$ pairing symmetry~\cite{wollman,shen}
indicate that there might be a strong
interdependence between the occurrence of high
transition temperatures and antiferromagnetic spin
fluctuations. Furthermore, the interesting dip
structures in ARPES spectra of
Bi$_{2}$Sr$_{2}$CaCu$_{2}$O$_{8+\delta}$ (Bi2212)
found for the first time by Dessau
{\it et al.}~\cite{dessau2} and confirmed by other
measurements~\cite{shen,hwu,randeria} is currently
debated and could be a further fingerprint of
the pairing mechanism.

{}From a theoretical point of view the two
 dimensional one-band Hubbard Hamiltonian
serves as a basic
model to describe the highly correlated electrons
within the $CuO_{2}$ planes. The first important
steps to investigate the superconducting
properties of the cuprates within this approach was done
simultaneously by Monthoux {\it et al.}~\cite{mont}
and Pao {\it et al.}~\cite{pao}, who found a
$d_{x^2-y^2}$ pairing state for low temperatures
 within a pertubative {\it ansatz}, namely the
fluctuation exchange approximation (FLEX).
Despite these interesting results the superconducting
excitation spectrum and in particular its doping
dependence are far from being understood.
Moreover,
it is still unclear if this microscopic approach
can explain the observed {\it optimal doping}
concentration in the High-T$_c$ superconductors.

In this article we present new results for the
doping dependence of the superconducting state of the
Hubbard Hamiltonian. We discuss the resulting
phase diagram and find that T$_c$ becomes maximal for a
doping concentration of $13 \; \%$. By determining
the excitation spectrum directly on the real frequency
axis we demonstrate that shadow bands as proposed
 by Kampf and Schrieffer~\cite{kampf} for the normal phase
are also present below $T_c$. We show that they
have an important impact on the pairing mechanism, on $T_c$
and on the gap function $\Delta ({\bf k},\omega)$.
In addition, they cause observable effects in
the spectral density of states and are responsible
for fine structures in ARPES experiments.

Our theory is based on a strong coupling Eliashberg
approach for the one-band Hubbard Hamiltonian
with nearest neighbor hopping integral $t=0.25 \, {\rm eV}$,
 bare  dispersion
$\varepsilon_0 ({\bf k})=(-2t[\cos(k_x)+\cos(k_y)]-\mu$)
with chemical potential $\mu$ and local Coulomb
repulsion $U=4t$. The superconducting state is treated
in the Nambu formalism, where the Greens function
$\hat G({\bf k},\omega)$ is a $2 \times 2$ matrix that can
be expanded in terms of Pauli matrices. The
corresponding expansion coefficients of the diagonal
electronic self energy are
$\omega (1-Z({\bf k},\omega))$ and $\chi({\bf k},\omega)$,
whereas
$\phi({\bf k},\omega) =
\Delta ({\bf k},\omega) Z({\bf k},\omega)$ is the
 coefficient of the anomalous
off-diagonal self energy that vanishes above $T_c$
and $\Delta ({\bf k},\omega)$ is the gap function.
These three functions were calculated self consistently
by using a newly developed numerical method for
treating the FLEX equations directly on the real frequency
axis. The calculations were performed on a
($64 \times 64$) square lattice in momentum space and by
using 4096 equally spaced energy points in the
interval $[-30 \;t,30 \;t]$ leading to a low energy
resolution of $0.014 \; t \approx 4$ meV~\cite{details}.

In the Fig. 1(a) we present the doping dependence of
 $T_c$, where $x$ is the doping concentration and
$n=1-x$ the occupation number per site. Note that the
order parameter was found to have a d$_{x^2-y^2}$
symmetry for all doping values. For large $x$, $T_c$
increases up to $T_c = 79 \; K$ at $x = 0.13$, whereas
below the {\it optimal doping} $T_c$ starts to decrease
for $x < 0.13$. Consequently, three different
doping regimes with qualitatively different behavior
occur within the Hubbard model: The {\em underdoped}
($x < 0.13$), the {\em optimally doped} and the
{\em overdoped} ($x > 0.13$) compounds. In
the following we will demonstrate $(i)$ that the
{\em optimally doped} systems are characterized by a
constructive interference of superconducting and
antiferromagnetic excitations, resulting from the
comparable time scales of both phenomena, $(ii)$
that the suppression of $T_c$ in the {\em overdoped}
systems is due to the increasing weakening of the
 effective pairing interaction, and $(iii)$ that in the
{\em underdoped} systems the dynamical character
 of the spin fluctuations becomes less pronounced causing
$T_c$ to decrease. Here the rather stiff arrangement
of antiferromagnetically correlated spins and their
interplay with the $d$-wave pairing state enhances
the formation of shadow states, i.e. states resulting
from the coupling of wave vectors ${\bf k}$ at the FS
and its shadow at ${\bf k+Q}$
(${\bf Q}= (\pi,\pi)$)~\cite{kampf}.

Important information about the origin of the doping
dependence of $T_c$ and the {\it optimal doping} can
be obtained from the effective interaction
$V_s ({\bf k},\omega)$ below $T_c$~\cite{mont}.
In Fig. 1 (b)
we compare $V_s ({\bf k},\omega)$ with
$V_n ({\bf k},\omega)$ that was obtained at the
same temperature,
but with $\phi \equiv 0$, for $\omega = 0$ and ${\bf k}$
points near ${\bf Q}$ where $V_s ({\bf k},0)$
is maximal. For {\it non-optimal doping} we find that
the interaction and in turn the
antiferromagnetic correlations increase only slightly
below $T_c$, whereas for compounds close to $x = 0.13$
an enhancement up to $40 \; \%$ occurs with respect
to the normal state. It is interesting that this happens
although our calculation yields that $V_s ({\bf k},\omega)$
increases continuously for decreasing doping even
below the {\it optimal doping}. The enhancement of
the pairing interaction is consistent with the interesting
doping dependence of $\Delta_{0} \equiv
 \Delta (T=0)=\phi (T=0)/Z (T=0)$~\cite{gap}.
The superconducting gap has
a maximum slightly above the {\it optimal doping}
and decreases sharply in the {\em underdoped}
compounds,
because the quasi particle scattering increases
 for smaller doping due to the strong antiferromagnetic
correlations via $Z$, whereas $\phi$ does not
change significantly for $x \leq 0.13$. In addition the
$2\Delta_{0}$ curve intersects the $\tau^{-1}$
curve at $x=0.13$, where  $\tau$ is the lifetime of the quasi
particle, and $\tau^{-1}\equiv \; {\rm Im}\;\Sigma
({\bf k},0)$ with FS-momentum ${\bf k}$~\cite{wsf} and self
energy $\Sigma ({\bf k},0)=\omega (1-Z({\bf k},\omega))
+\chi({\bf k},\omega)$ at $T_c$. In Fig. 1 (b) one
sees that $\tau^{-1}$ increases monotonously with
decreasing doping and saturates for small and large values
of $x$ as observed in transport measurements~\cite{batlogg}.
Therefore the occurrence of an {\it optimal doping}
in our results is related to an
interference of the typical lifetime of the Cooper
 pairs $\sim \Delta^{-1}_{0}$ and the lifetime $\tau$
of the quasi particles: First $T_c$ increases for
decreasing doping, since $V_s ({\bf k},\omega)$
increases
and since $\tau$ is large enough to guarantee an
 effective Cooper pair formation. Then below the
{\it optimal doping}, $T_c$ decreases again even
so $V_s ({\bf k},\omega)$ continues to increase,
since $\tau$
becomes too small and there are no well defined
quasi particles during the pairing process. Notice,
that there
are also remarkable variations of the ratio
$2 \Delta_{0} / k_B T_c$ upon doping shown in Fig. 1(c).
The
deviations of $2 \Delta_{0} / k_B T_c$ from the BCS
value demonstrate the qualitative new changes of the
superconducting state for decreasing temperatures,
which are mostly pronounced at the {\it optimal doping},
where a strong increase of the pairing interaction occurs.

The interdependence of dynamical antiferromagnetism
and superconductivity leads not only to an
{\it optimal doping}, but also to a different behavior of
the gap function
$\Delta ({\bf k},\omega)$ in {\em overdoped} and
{\em underdoped} compounds. In Fig. 2 we plot
$\Delta ({\bf k},\omega)$ for different frequencies along
a path in the  Brillouin zone as indicated
in the insets. By comparing our results with the
{\it ansatz} $\Delta_d \sim \cos (k_x) - \cos (k_y)$ one
sees deviations from the simple $d$-wave behavior.
These are mostly pronounced for small frequencies and
in particular for {\em underdoped} systems, where
the gap function is enhanced and largest at the FS and
at its shadow. By transforming our results for
$\Delta ({\bf k},\omega)$ to real coordinate space,
we find
that this corresponds to a pairing dominated by
nearest-neighbor interactions in the case of $x=0.16$.
However, the appearance of higher harmonics in
$\Delta ({\bf k},\omega)$ for $x=0.09$ indicates that
pairing processes between more distant
antiferromagnetically correlated spins come into play.
This results
from the increasing stiffness of the nearest-neighbor
spin correlations, which do not contribute any longer
as strongly to the spin fluctuation mediated pairing
 interaction.

Further insight into the effect of the short range
antiferromagnetic order on the excitations can be
obtained by investigating the frequency dependence
of the quasi particle decay. This is
dominated by the coupling of states at the FS and its
shadow. By inverting the matrix Greens function
$\hat G({\bf k},\omega)$ one can define an effective
electronic self energy that includes the off-diagonal
contributions and is equal to $\Sigma ({\bf k},\omega)$
in the normal state~\cite{bick}:
\begin{equation}
\Sigma_{\phi} ({\bf k},\omega)=\omega
(1-Z({\bf k},\omega))+\chi({\bf k},\omega)
+\frac{(\phi({\bf k},\omega))^2}{\omega Z({\bf k},\omega)
+ (\varepsilon_0 ({\bf k})-\mu)+
\chi({\bf k},\omega)}\;.
\label{epl1}
\end{equation}
In Fig. 3 we compare the imaginary part of
$\Sigma_{\phi} ({\bf k},\omega)$ at the FS and at its
shadow for an {\em overdoped} and an {\em underdoped}
compound above and below $T_c$. In the normal
state we find for $x=0.16$ that
${\rm Im}\;\Sigma_{\phi} ({\bf k},\omega)$ is linear in $\omega$
down to energies $\omega \approx 8 \;
{\rm meV} \approx 80\;K$, which is consistent with t
he marginal
Fermi-liquid (MFL) theory~\cite{VLS89}. For $x=0.09$
we find besides an overall increase of the scattering
rates a minimum (III) at the shadow at the FS that is
a precursor of the singular behavior of
$\Sigma_{\phi} ({\bf k},\omega)$ in the antiferromagnetic
state. Thus the {\it underdoped} compounds are
clearly not in agreement with the properties of an MFL.
Below $T_c$ and for $x=0.16$ a sharp minimum (I)
becomes visible at the Fermi energy ($\omega=0$)
when ${\bf k}$ is close to the FS which is the analogue of
 the corresponding $\delta$-function in the
BCS theory~\cite{bcs}. Thus minimum I shifts well
below the Fermi energy when ${\bf k}$ approaches
the shadow
of the FS. Furthermore, its intensity decreases and
becomes almost equal to the additional minimum (II) at
$\omega \approx + 50$ meV in Fig 3(a) and (b). Note
that minimum II vanishes for ${\bf k}$ points away from
the FS and its shadow (not shown) and is purely caused
by the opening of the superconducting gap and a
related shift of spectral weight in the effective interaction.
Hence, this dip reflects the enhancement of
the antiferromagnetic coupling below $T_c$ as already
discussed in the context of Fig. 1(b). A similar
feedback effect between spin fluctuations and $d$-wave
pairing state occurs in the {\em underdoped}
system. Here, the strongly enhanced scattering rates and
the smaller superconducting gap lead to a rather small
minimum I at the FS with respect to $x=0.16$ and to a
much weaker minimum II above the Fermi energy, whose
energy shifts to the Fermi level for decreasing doping.
However and more interestingly, the remarkable
enhancement of minimum III shows that it is not a
simple superposition of the precursor of the
antiferromagnetic singularity and the superconducting
singularity, but caused by a real interdependence
between both phenomena.

To demonstrate that the interplay between dynamical
antiferromagnetism that causes shadow states in the
cuprates and superconductivity  has not only
consequences for the doping dependence of $T_c$,
but can also
be observed in ARPES experiments we extended
previous results for the spectral density of states
$\rho ({\bf k},\omega)$~\cite{supra1} to analyze what
happens when ${\bf k}$ crosses the shadow of the FS.
 In Fig. 4 we plot $\rho ({\bf k},\omega)$ for ${\bf k}$
close to the shadow of the FS in the neighborhood of
the ($\pi,0$) point. For the {\em overdoped} system
one sees that the maximum II in
$\Sigma_{\phi} ({\bf k},\omega)$ of Fig. 3 leads to a
corresponding dip in
$\rho ({\bf k},\omega)$ at $\omega \approx + 60$ meV,
where spectral weight is suppressed with respect to
the normal state. This dip-structure agrees well with
the experimental finding of an anomalous dip
structure at the shadow of the FS which is mostly
pronounced near the ($\pi,0$) point with energy
$\omega_{exp} \approx 70$ meV~\cite{dessau2,topo}.
In the normal state of the {\it underdoped} system the
small satellites below the Fermi level, which are separated
from the main band above the Fermi energy, are
the shadow states. They become visible below $x=0.13$
and increase with decreasing doping. Below $T_c$ the
dip structure as discussed for $x=0.16$ shifts to smaller
energies leading to a transfer of spectral weight
below the Fermi level~\cite{supra1}. Note that a first
experimental indication for this anomalous transfer
of spectral weight was recently obtained by
 Dessau {\it et al.}~\cite{dessau2}.

Summarizing, we presented new results for the
superconducting properties and the phase diagram
of the 2D
Hubbard model (Fig. 5). In particular, we obtained
an {\it optimal doping} as a result of the
constructive interference between magnetic
correlations and $d$-wave pairing state. Furthermore, we
demonstrated for the first time that shadow states
exists below $T_c$ which are related to dip structures
and to an anomalous transfer of spectral weight at
the shadow of the FS in the spectral density of states.
These were observed in photoemission experiments
indicating the importance of spin fluctuations for the
formation of superconductivity in the cuprates.
%
%
%
%

%
\begin{figure}
\caption{Doping dependence of the superconducting
state: (a) $T_c$ obtained from $\Delta(T=T_{c})=0$.
(b) Enhancement of the effective interaction
max($V_s ({\bf k},0))/$max$(V_n ({\bf k},0)$), $2 \Delta_{0}$
and inverse quasi particle lifetime $\tau^{-1}$. (c)
 Ratio $2 \Delta_{0}/ k_B T_c$.}
\label{fig1}
\end{figure}
\begin{figure}
\caption{Superconducting gap function
$\Delta ({\bf k},\omega)$ for different frequencies and along
different paths in the Brillouin zone (see inset).
Note that $\Delta ({\bf k},\omega)$ is normalized
by $\Delta ({\bf k}=(\pi,0),\omega)$ and that for
comparison we plot the simple $d$-wave gap function
$\Delta_d \sim \cos (k_x) - \cos (k_y)$.}
\label{fig2}
\end{figure}
\begin{figure}
\caption{ ${\rm Im}\;\Sigma_{\phi} ({\bf k},\omega)$
at the Fermi surface and at its shadow. Note, the
results $T=T_c$ refer to the normal state.}
\label{fig3}
\end{figure}
\begin{figure}
\caption{Density of states $\rho ({\bf k},\omega)$
along a path in the Brillouin zone that
crosses the shadow of the Fermi surface above
and below $T_c$.}
\label{fig4}
\end{figure}
\begin{figure}
\caption{ Schematic phase diagram of the 2D
Hubbard model, where the antiferromagnetic phase is only
plotted for illustration although we get no magnetic
phase transition in two dimension.}
\label{fig5}
\end{figure}
%
%

\begin{references}
\bibitem{details}
Details will be published elsewhere. Note, a version of
the Eliashberg equations on the
imaginary axis is given in Ref.~\onlinecite{mont}.
\bibitem{gap}
For the sake of comparison with the experiments
we estimated $2 \Delta_{0}$ from the distance between
the superconducting pileups in the momentum
averaged density of states $\rho(\omega)$.
\bibitem{wsf}
Note, we find a similar behavior of the quasi particle
lifetimes at the shadow of the FS which is
also important for the density of states at the Fermi energy.
\end{references}
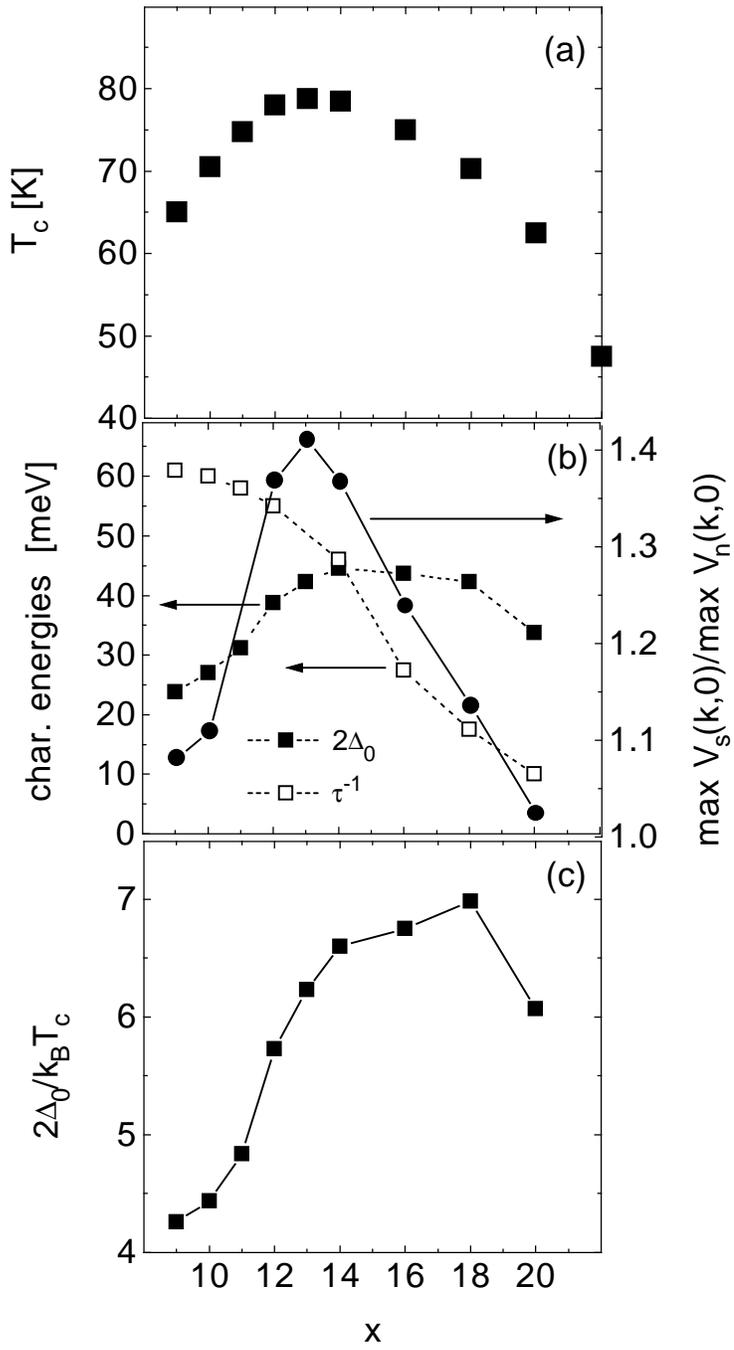
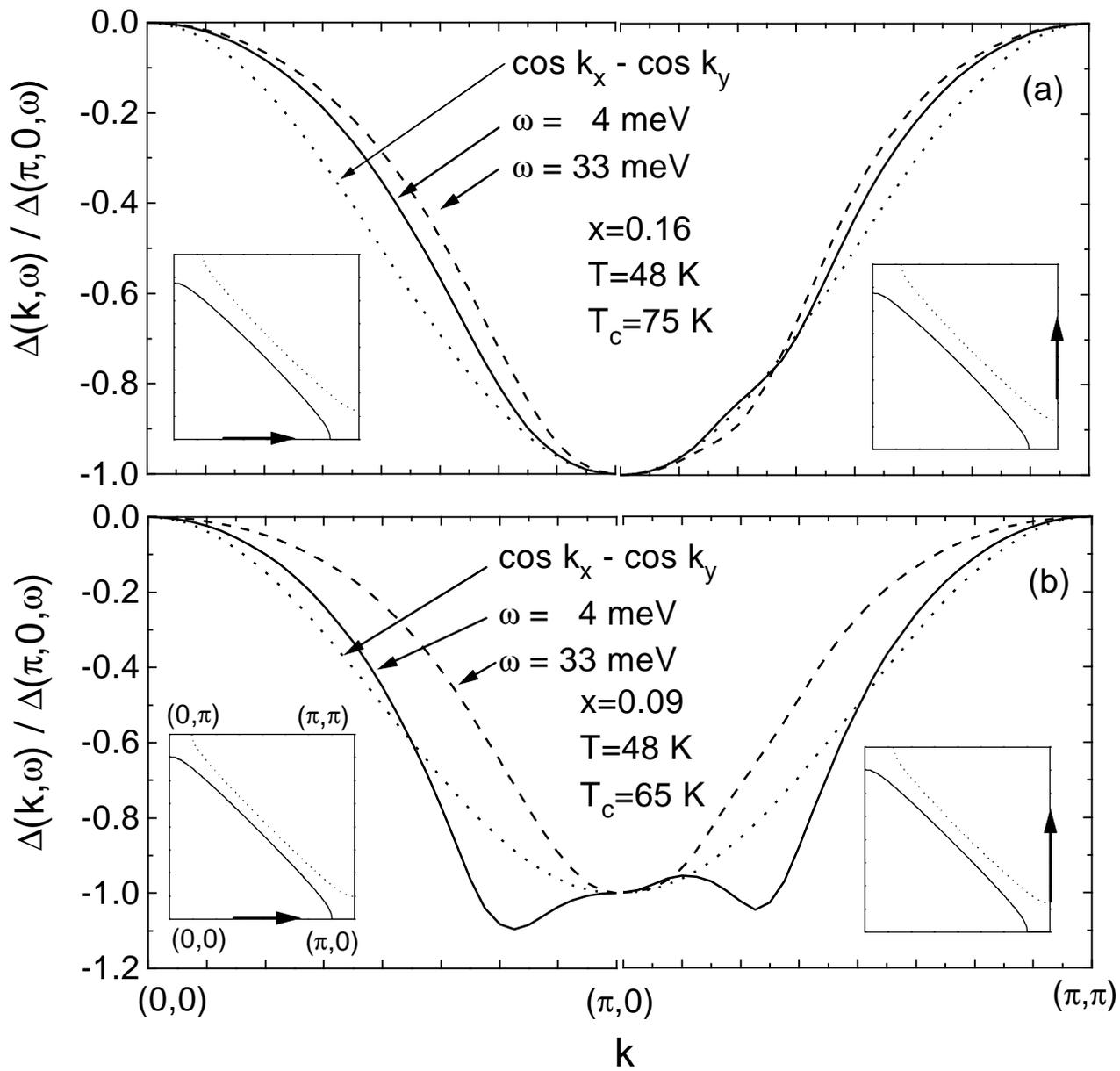
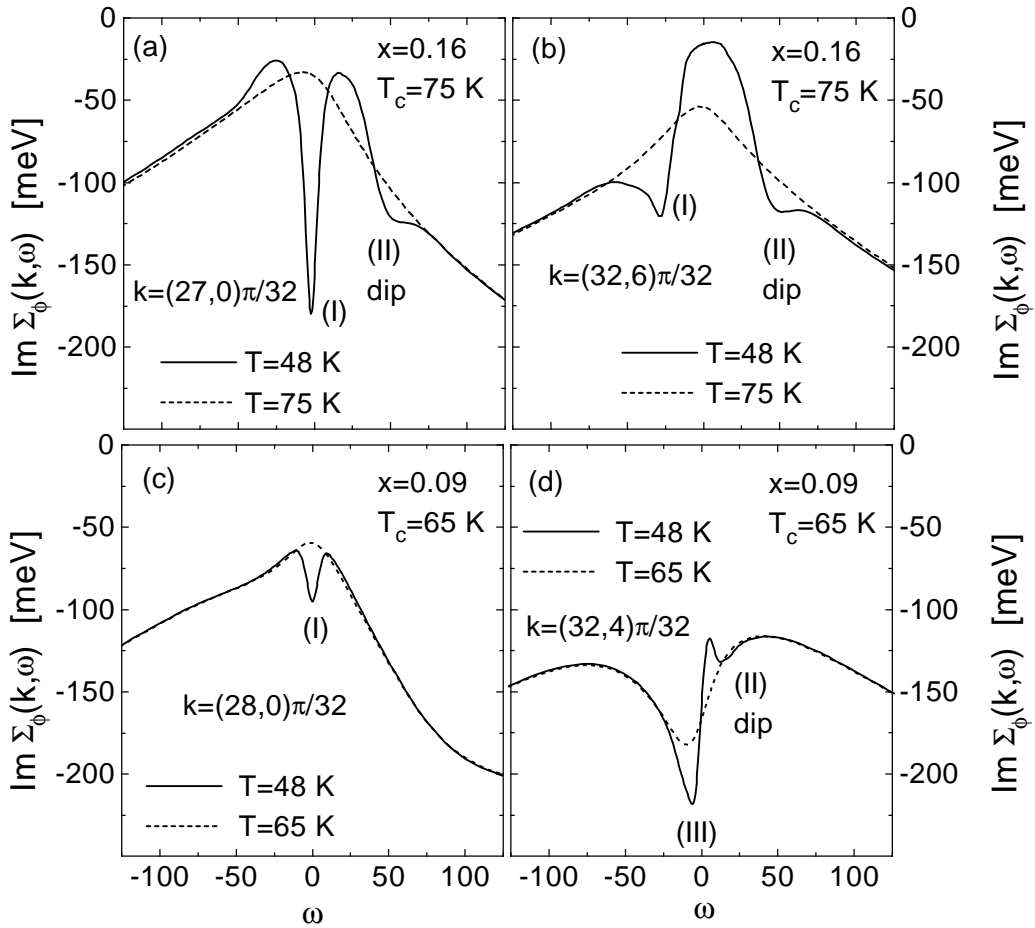
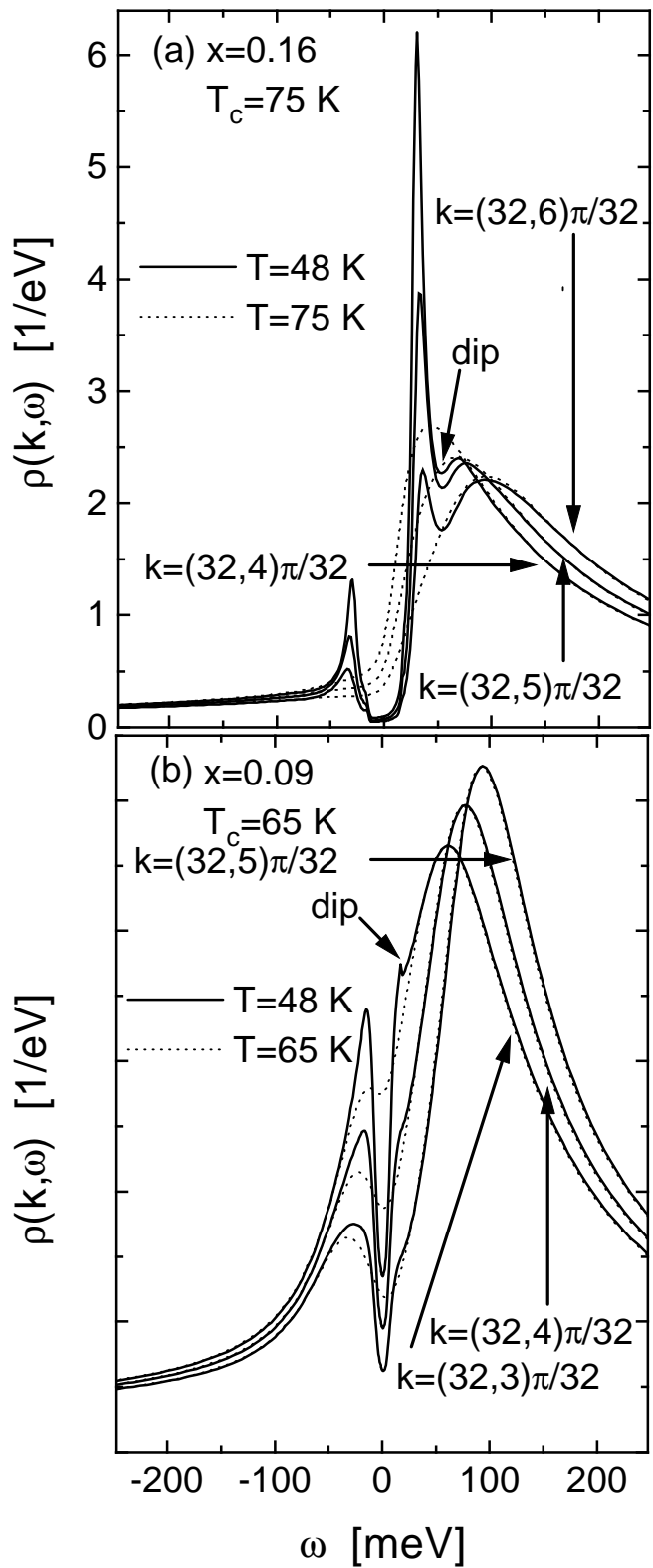
\end{document}